\documentclass[prl,twocolumn,floatfix,amssymb,amsmath,nobibnotes,superscriptaddress,showpacs]{revtex4-1}

\usepackage{color}
\usepackage{graphicx} 
\usepackage{subfigure}
\usepackage{dcolumn}  
\usepackage{amsmath,amssymb}
\usepackage{epstopdf}

\def\Eq#1{Eq.~(\ref{#1})}
\def\e{\epsilon}

\def\sc{_{\rm sc}}

\def\T{\beta}

\def\beal{\begin{align}}
\def\eal{\end{align}}

\def\bea{\begin{eqnarray}}
\def\eea{\end{eqnarray}}
\def\ben{\begin{equation}}
\def\een{\end{equation}}
\def\benu{\begin{enumerate}}
\def\enu{\end{enumerate}}

\def\bei{\begin{itemize}}
\def\eei{\end{itemize}}
\def\beit{\begin{itemize}}
\def\eit{\end{itemize}}
\def\benu{\begin{enumerate}}
\def\enu{\end{enumerate}}

\def\sss{\scriptscriptstyle\rm}

\def\1var{(\bx_1...\bx\N)}

\def\br{{\bf r}}

\def\bx{{x}}

\def\s{_{\sss S}}
\def\xc{_{\sss XC}}

\def\N{_{\sss N}}
\def\H{_{\sss H}}

\def\ee{_{\rm ee}}

\begin{document}


\title{Bypassing the malfunction junction in warm dense matter simulations}

\author{Attila Cangi}
\affiliation{Max Planck Institute of Microstructure Physics, 
Weinberg 2, 06120 Halle (Saale), Germany}
\email[]{acangi@mpi-halle.mpg.de}

\author{Aurora Pribram-Jones}
\affiliation{Department of Chemistry,
University of California, Irvine, CA 92697-2025, USA}

\pacs{71.15.Mb,31.15.E-}

\date{\today}

\begin{abstract}  
Simulation of warm dense matter requires computational
methods that capture both quantum and classical behavior efficiently under
high-temperature, high-density conditions.  Currently, density functional
theory molecular dynamics is used to model electrons and ions, but this
method's computational cost skyrockets as temperatures and densities
increase.  We propose finite-temperature potential functional theory as an 
in-principle-exact alternative that suffers no such drawback. We derive an orbital-free free energy approximation through a coupling-constant formalism.  Our density approximation and its
associated free energy approximation demonstrate the method's accuracy and
efficiency. 
\end{abstract}

\maketitle

Warm dense matter (WDM) is a highly energetic phase of matter with
characteristics of both solids and plasmas\cite{GDRT14}. The high temperatures
and pressures necessary for creation of WDM are present in the centers of
giant planets and on the path to ignition of inertial confinement fusion
capsules\cite{A04,HEDP03}. The high cost of experiments in this region of
phase space has led to renewed interest and great progress in its theoretical
treatment\cite{MD06,GBBC12,STVM00}. Since both quantum and classical effects
are crucial to accurate WDM simulations\cite{KD09}, density functional theory
molecular dynamics has been used with increasing frequency\cite{HRD08}.  This
method relies on Kohn-Sham density functional theory (KS-DFT), which simplifies
solving the interacting many-body problem of interest by mapping it onto a 
non-interacting system\cite{HK64,KS65}. While the
agreement between these calculations and experimental results is
excellent\cite{KRDM08,RMCH10}, the calculations are still incredibly expensive\cite{MH00,MH09}.  
The computational bottleneck in these calculations is the cost of solving the KS
equations, a step that becomes increasingly expensive as temperatures rise and high-energy states become fractionally occupied. In fact, the computational cost exhibits nearly exponential scaling with temperature
due to the KS cycle including a high number of states at the temperatures
of WDM\cite{KSCD14}.

A solution to this problem is orbital-free DFT\cite{WC00}, which avoids
this costly step by using non-interacting kinetic energy approximations that
depend directly on the electronic density.  Because the kinetic energy is such
a large fraction of the total energy, however, these approximations must be
highly accurate to be of practical use.  Though much progress has recently
been made for WDM\cite{KST12,KCST13,SD13}, approximations for thermal
ensembles are complicated by temperature effects. The KS kentropy, the
free energy consisting of the non-interacting kinetic energy and entropy,
must be approximated directly, greatly complicating the production of
useful and efficient approximations.
\begin{figure}[t]%
\begin{center}
\includegraphics[width=8.25cm]{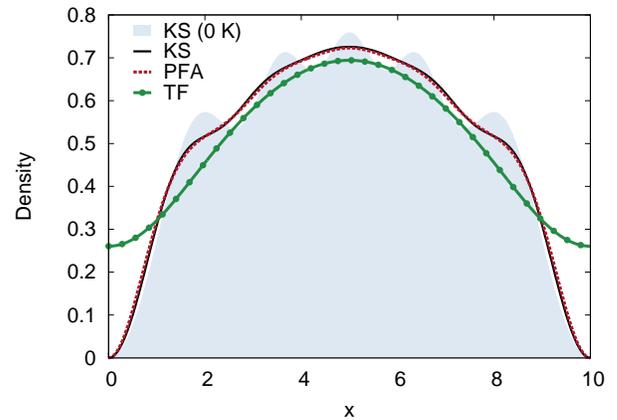}
\end{center}
\caption{
Shortcomings of the TF approximation in the WDM regime: 
Total density of five particles in the potential 
$v(x) = -2 \sin^2(\pi x/10)$ within a box (of size 10 a.u.) 
at $\Lambda=1/(\beta\mu)=0.93$.
Compare the exact density (solid black curve) 
with our PFA (dashed red curve) derived in Eq.~(\ref{n.finT}),
which is basically on top of the exact result. 
On the other hand, the TF approximation (dotted green curve) captures the 
general qualitative features, but completely misses the quantum oscillations. 
We also show the corresponding exact density at zero temperature 
(light blue shaded area), with its pronounced oscillations that smooth as temperatures rise.}
\label{f:n}%
\end{figure}

At zero temperature, potential functional theory (PFT) is a promising,
alternative approach to  the electronic structure problem\cite{CLEB11,CGB13}.
It is also orbital-free, but skirts the troublesome issue of separately
approximating the KS kinetic energy.  PFT's coupling-constant formalism
automatically generates a highly accurate kinetic energy potential functional
approximation (PFA) for any density PFA\cite{CLEB11}.  In this way, one needs
only to find a sufficiently accurate density approximation, as has been
demonstrated for simple model systems\cite{CLEB10}. 
Approximations to the non-interacting density have been derived in various
semiclassical\cite{ELCB08,CLEB10,RBK10} and stochastic approaches\cite{BNR13}.
It has also been shown that, at zero temperature, PFT generates leading
corrections to local approximations\cite{CLEB10}, which become exact in
the well-known Lieb limit\cite{LS73}. Finite-temperature Thomas-Fermi theory\cite{T27,F28} 
has been shown to become relatively exact for non-zero temperatures under scaling similar 
to that used by Lieb\cite{NT80}. 
In this way, our method provides a pathway to
systematic improvements in approximations to the kentropy, something
generally missing from DFT approaches.

The particular scaling conditions under which TF becomes exact for all
temperatures is related to the breakdown of purely quantum or purely classical
behavior as both temperatures and particle numbers increase\cite{GDRT14}.  
The importance of both these effects in the WDM regime underlies its theoretical
complexity\cite{PPGB14}, so it is useful to represent the influences of
temperature and density with a single parameter,
an electron degeneracy parameter defined by $\Lambda = 1/(\beta\mu)$, 
which depends on the system temperature $1/\beta$ and temperature-dependent chemical potential $\mu$.  
Then, the WDM regime can be defined as where $\Lambda\approx 1$.  At these conditions, 
ground-state DFT is hugely expensive, while traditional plasma methods miss critical
electronic structure features.  In Fig. \ref{f:n}, we show the density
oscillations still present at WDM conditions that are neglected by the
smooth, classical TF approximation, but are captured by our method.

In this work, we 
(i) derive PFT for thermal ensembles, 
(ii) give an explicit equation for the kentropy relying solely on the
temperature-dependent density,
(iii) derive and implement a highly accurate density approximation in one dimension 
to illustrate our general result that
contains the leading correction TF theory at finite
temperature, and
(iv) perform (orbital-free) PFT calculations in the WDM regime.
Our method generates highly accurate density and kentropy approximations,  
skirts the need for separate kentropy approximations, provides a roadmap
for systematically improved approximations, and converges more quickly as temperatures 
increase while maintaining accuracy at low temperatures. 
At the same time, it bridges low and high temperature methods, and so is uniquely suited to WDM.

At non-zero temperature, the energy is replaced by the grand canonical potential 
as the quantity of interest\cite{M65,E10}. The grand canonical Hamiltonian is written
\ben\label{gcop}
\hat{\Omega} = \hat{H}  - \frac{1}{\beta} \hat{S} - \mu \hat{N},
\een
where $\hat{H}$, $\hat{S}$, and $\hat{N}$ are the Hamiltonian, entropy, 
and particle-number operators.  In electronic structure theory, 
we typically deal with non-relativistic, interacting electrons, most commonly within 
the Born-Oppenheimer approximation.
The electronic Hamiltonian (in atomic units here and thereafter) reads 
\ben
\hat H = \hat T + \hat V\ee + \hat V\,,
\een
where  $\hat T$ denotes the kinetic energy operator,
$\hat V\ee$ the interelectronic repulsion, 
and $v(\br)$ denotes the static external potential in which the electrons move.
(We suppress the spin of the electron for simplicity of notation.)
The grand canonical potential can be written in terms of potential functionals as follows:
\ben
\Omega^{\sss \T}_{v-\mu} = F^{\sss \T}[v] + \int d^3r~n^{\sss \T}[v]({\bf r}) (v({\bf r})-\mu)
\een
with
$F^{\sss \T}[v] =  F^{\sss \T} [\hat{\Gamma}^0_{v-\mu}]  = 
T[\hat{\Gamma}^0_{v-\mu}] + V_{ee}[\hat{\Gamma}^0_{v-\mu}] 
- \frac{1}{\beta} S[\hat{\Gamma}^0_{v-\mu}]$ 
denoting the universal functional\cite{PPFS11}. 

In practice, approximating this direct expression would require two separate approximate potential functionals, 
one for the universal finite-temperature functional and one for the density:
\ben
\breve{\Omega}^{\sss \T,{\sss dir}}_{v-\mu} 
= \breve{F}^{\sss \T}[v] + \int d^3r~\breve{n}^{\sss \T}[v]({\bf r}) (v({\bf r})-\mu).
\een
However, we can generate an approximation (denoted by a breve above the approximated quantity) 
to the universal functional that corresponds to any density approximation.  In analogy 
to the zero-temperature case\cite{CLEB11}, we introduce a coupling 
constant $\lambda$ in the one-body potential,
$v^\lambda(\br)=(1-\lambda)v_0(\br)+\lambda v(\br)$,
where $v_0$ is some reference potential. Via the Hellmann-Feynman theorem 
we can then rewrite the grand potential,
\ben
\Omega^{\sss \T}_{v-\mu}=\Omega^{\sss \T}_0+\int_0^1 d\lambda\int d^3r~n^{\sss \T}[v^\lambda]({\bf r})\Delta v(\br),
\een
where $\Delta v(\br)=v(\br)-v_0(\br)$.  
Setting $v_0=0$ and defining $\bar{n}^{\sss \T}[v](\br)=\int_0^1 d\lambda\, n^{\sss \T}[v^\lambda](\br)$, 
we can now write the finite-temperature universal functional in terms of the density written 
as a potential functional:
\ben\label{Fcc}
F^{\sss \T,cc}_{n^{\sss \T}}[v]=\int d^3r~\left\{\bar{n}^{\sss \T}[v](\br)-n^{\sss \T}[v](\br)\right\}v(\br).
\een  
This defines an approximate functional, $\breve{F}^{\sss \T,cc}_{\breve{n}^{\sss \T}}[v]$, 
corresponding to the chosen density approximation $\breve{n}^{\sss \T}$ and
is the generalization of PFT to thermal ensembles.

Since practical use of this formula as written would require sufficiently
accurate approximations to the interacting electron density 
(which are likely unavailable), we instead apply it 
to the non-interacting electrons of the KS system. 
In DFT, the KS system is a clever way of approximating the exact $F^{\sss \T}$ 
by mapping the interacting system to a non-interacting system with the same electronic density 
and temperature.  This determines the one-body KS potential and corresponding chemical potential.  
Through this mapping, the non-interacting, finite-temperature universal density functional 
can be defined\cite{PPFS11}
\ben
F_{\sss s}^{\sss \T}[n] :=  \min_{\hat{\Gamma}\to n} K^{\sss \T}[\hat{\Gamma}] 
= K^{\sss \T}[\hat{\Gamma}_{\sss s}^{\sss \T}[n]]=K_{\sss s}^{\sss \T}[n],
\een
which generates the KS equations and, through their solutions, the KS orbitals. 
Slater determinants of the KS single-particle orbitals are the KS wavefunctions. 
The orbitals are implicit functionals of the density via the KS equations, 
and the average density can be constructed by Fermi-weighted summing of the orbitals.  
Solution of these equations at every time-step is the most costly step of DFT molecular dynamics. 

The KS potential is defined\cite{CLEB11,CGB13}
\ben\label{vspft}
v\s(\br) = v(\br) + \tilde v\H[n\s^{\sss \T}[v\s]](\br)
+ \tilde v\xc[n\s^{\sss \T}[v\s]](\br)\,,
\een
where, in contrast to KS-DFT, the density is posed as a \emph{potential} functional,
and tildes distinguish density functionals from potential functionals.  
All many-body interactions among the electrons are captured in the usual KS-DFT sense, 
via the (traditionally defined) Hartree and XC potentials\cite{DG90}.
The difference from a usual KS-DFT calculation is that 
\Eq{vspft} in conjunction with an approximation to the non-interacting density 
\emph{bypasses} the hugely expensive iterative solution of the KS equations for WDM.
Choosing a \emph{potential} functional approximation  
to the non-interacting density automatically generates an approximated KS potential, 
as illustrated in the Supplemental Materials.
Once the self-consistent KS potential is determined, 
the KS kentropy is computed from
\ben\label{Kscc}
K_{{\sss S}, n\s^{\sss \T}}^{\sss \T,{\rm cc}}[v\s]
=\int d^3r~\left\{\bar{n}\s^{\sss \T}(\br)-n\s^{\sss \T}[v\s](\br)\right\}v\s(\br)\,, 
\een 
which is the analog of \Eq{Fcc} for KS electrons.
Again, \Eq{Kscc} defines a coupling-constant approximation, 
$\breve{K}_{{\sss S}, \breve{n}\s^{\sss \T}}^{\sss \T,{\rm cc}}[v\s]$, 
when evaluated on any chosen approximation to the non-interacting density $\breve{n}\s^{\sss \T}$.
Finally, the grand potential expressed in terms of KS quantities\cite{PPFS11},
\begin{align}
\begin{split}
\Omega^{\sss \T}_{v-\text{\scriptsize{$\mu$}}} 
&= K_{{\sss S}, n\s^{\sss \T}}^{\sss \T,{\rm cc}}[v\s] 
+ U[n\s^{\sss \T}[v\s]] 
+ {\cal F}^{\sss \T}_{xc}[n\s^{\sss \T}[v\s]] \\
&+ \int d^3r~n^{\sss \T}[v\s]({\bf r}) \left( v({\bf r})-\mu \right)\;,
\end{split}
\end{align}
can be evaluated via \Eq{Kscc}. 
Through this result, we leverage the body of time-proven XC approximations 
and eliminate the need to construct separate approximations to the KS kentropy 
for use in orbital-free (and thereby computationally inexpensive) schemes.
Only an approximation to the non-interacting density is required. 
This means that a general, systematic, non-empirical route 
to improved kentropy approximations is now available.

In principle, a possible starting point for deriving an approximation 
to the non-interacting density at finite temperature is the semiclassical propagator
\ben\label{ftprop}
G\sc^{\sss \T}(\br,\br';\alpha) = G\sc^0(\br,\br';\alpha)\,f^{\sss \T}(\alpha)\,, 
\een
which can be written as a convolution of the zero-temperature propagator 
$G\sc^0(\br,\br';\alpha)$ with the factor 
$f^{\sss \T}(\alpha) = \pi \alpha/[\beta\sin(\pi\alpha/\beta)]$ 
carrying all temperature dependence, and $\alpha$ is a complex variable\cite{BBD85}.
From the propagator, we extract the density via an inverse Laplace transformation\cite{G60}
\ben\label{nsc_3d}
\breve{n}\s^{\sss \T}[v\s](\br) = \lim_{\br'\to\br} 
\frac{1}{2\pi i}\int\limits_{\eta-\infty}^{\eta+\infty} d\alpha\ \frac{e^{\mu\alpha}}{\alpha}\, 
G\sc^{\sss \T}[v\s](\br,\br';\alpha)\ . 
\een 

To illustrate the significance of our main result in Eq.~(\ref{Kscc}), 
we consider a simple, yet useful, numerical demonstration: 
Non-interacting, spinless fermions in an arbitrary 
potential $v(x)$ confined to a box of size $L$ 
obeying vanishing Dirichlet boundary conditions. 
(In a practical realization, this would
be the self-consistent KS potential of the given many-body problem.)
Recently, a highly accurate PFA to the density was derived for this model
using the path integral formalism and semiclassical techniques\cite{CSB14}. 
Here we extend this result to finite temperature via \Eq{ftprop}
and obtain: 
\ben\label{n.finT}
\breve{n}\s^{\sss \T}(x)= 
\lim_{x'\to x} \sum_{\lambda=1}^4\sum_{j=0}^\infty 
\breve{\gamma}\s^{\sss \T}(x,x';\lambda,j) \,,
\een
a PFA to the density at a given temperature and chemical potential, 
where
\ben
\breve{\gamma}\s^{\sss \T}(x,x';\lambda,j)=
\frac{\sin{\Theta_\mu^\lambda(x,x';j)}\textrm{csch}[\frac{\pi}{\beta} {\mathcal T}_\mu^\lambda(x,x';j)]}
{(-1)^{\lambda+1} \beta\sqrt{k_\mu(x)k_\mu(x')}} \ .
\een
Here we define generalized classical phases
$\Theta_\mu^1(x,x';j) = \theta_\mu^-(x,x') +2 j \theta_\mu(L)$,
$\Theta_\mu^2(x,x';j) = \theta_\mu^+(x,x') +2 j \theta_\mu(L)$,
$\Theta_\mu^3(x,x';j) = \theta_\mu^-(x,x') -2 (j+1) \theta_\mu(L)$,
$\Theta_\mu^4(x,x';j) = \theta_\mu^+(x,x') -2 (j+1) \theta_\mu(L)$
and generalized classical traveling times
${\mathcal T}_\mu^\lambda(x,x';j)=d \Theta_\mu^\lambda(x,x';j) /d\mu$.
Furthermore,
$\theta^\pm(x,x') = \theta(x)\pm\theta(x')$, where
$\theta_\mu(x)= \int_0^x dy\, k_\mu(y)$ and
$k_\mu(x)=\sqrt{2(\mu-v(x))}$ at a given chemical potential $\mu$,
which is determined by normalization of the density.

The physical interpretation of our result in \Eq{n.finT} is instructive:
For a given chemical potential there are infinitely many classical paths that 
contribute to the total density. The paths are classified into four primitives
(identified by $\lambda$) onto which an integral number of periods (labelled by $j$) 
is added.  
The first primitive is special, in the sense 
that it yields the TF density. 
All other primitives and additional 
periods carry phase information about reflections from the boundaries, producing 
quantum density oscillations that greatly improve upon the TF result\cite{CSB14}.
For a more details on the derivation and the physical interpretation,
we refer to Ref.~\cite{CSB14}.

Our result in Eq.~(\ref{n.finT}) can be evaluated numerically for a given temperature
by truncating the infinite sum at a conveniently chosen upper limit such that the 
result is convergent.  Importantly for WDM applications, the higher the temperature, 
the lower the upper limit required for convergence of the sum. 
In fact, in the WDM regime only the leading term ($j=1$) in the sum 
needs to be kept. While especially powerful at finite temperature, this might be a universal feature 
due to the semiclassical nature of our approximation. Similar results have also been 
recently found at zero temperature\cite{RB08,CSB14}.

However, the stationary phase
approximation used to derive Eq.~(\ref{n.finT}) 
yields the TF density at zero temperature as the leading term,
i.e., $\lim_{x'\to x} \breve{\gamma}\s^{\sss \T}(x,x';1,0)~=~k_\mu(x)/\pi=\breve{n}^0_{\sss TF}(x)$,
instead of the finite-temperature TF density 
$\breve{n}^{\sss \T}_{\sss TF}(x)=\sqrt{1/(2\pi^2\beta)}\,F(z)$,
where $F(z)~=~\int_0^\infty da\, \{\sqrt{a}[1+\exp(a-z)]\}^{-1}$ and $z=\beta\,k_{\mu}^2(x)/2$. 
We fix this problem with an ad-hoc correction and ensure the correct boundary conditions. 
To do so, we replace the density from the first primitive $\lim_{x'\to x} \breve{\gamma}\s^{\sss \T}(x,x';1,0)$ 
with a Gaussian interpolation of $\breve{n}^0_{\sss TF}(x)$ and 
$\breve{n}^{\sss \T}_{\sss TF}(x)$. 
In this way, we cope with the density approaching the high-temperature limit 
(under which TF theory becomes exact) differently 
in two distinct regions, the interior of the box and the edge regions near the walls.
These may be considered as two distinct boundary layers with different asymptotic expansions 
in the high-temperature limit. Note that the size of the boundary layers 
in the edge regions shrinks as the limit is approached.  
The Gaussian interpolation applied here is a crude version of the approach
used in boundary-layer theory to match two different approximations 
with different asymptotic behavior\cite{H13}.

In Fig.~\ref{f:n}, we plot a typical density of five particles in the WDM regime ($\Lambda\approx 1$) 
in the potential $v(x) = -2 \sin^2(\pi x/10)$ within a ten-unit box,
along with approximate densities. The black curve is the exact result, 
the red dashed curve is our approximation, 
and the green dotted curve is the TF density. In addition, the light-blue shaded area denotes 
the corresponding density at zero temperature. This figure demonstrates that quantum oscillations 
in the density persist in the WDM regime and that TF theory completely fails to capture them.
On the other hand, our PFA -- derived to include quantum effects -- is able to describe
them properly and is therefore highly accurate.
\begin{table}[ht]
\caption{Residual kentropy  
of five particles in the same potential as in Fig.~\ref{f:n}.
We list the error of the conventional TF approach and of our PFA (given in \Eq{Kscc.model}) 
far below and above where WDM is typically encountered.}
\label{t:K.L10D2m1}
\begin{ruledtabular}
\begin{tabular}{ d d d d d }
\multicolumn{1}{c}{$\Lambda$} &
\multicolumn{1}{c}{$K_{{\sss S},0}^{\sss \T}$} &  
\multicolumn{1}{c}{$\Delta K\s^{\sss \T}$} &  
\multicolumn{2}{c}{$\textrm{error}~\times~10^2$}\\
\hline
\multicolumn{3}{c}{} &
\multicolumn{1}{c}{TF} &
\multicolumn{1}{c}{PFA} \\
\cline{4-5}
0.16 &  3.94 & 0.462 & 6.39 & -0.32\\
0.31 &  3.87 & 0.461 & 7.16 & -0.28\\
0.47 &  3.76 & 0.459 & 7.91 & -0.31\\
0.62 &  3.64 & 0.456 & 8.39 & -0.29\\
0.78 &  3.50 & 0.452 & 8.61 & -0.30\\
0.93 &  3.34 & 0.448 & 8.65 & -0.37\\
1.09 &  3.16 & 0.444 & 8.58 & -0.50\\
1.40 &  2.77 & 0.435 & 8.22 & -0.87\\
1.71 &  2.36 & 0.425 & 7.69 & -1.27\\
2.02 &  1.92 & 0.414 & 7.13 & -1.61\\
2.48 &  1.25 & 0.396 & 6.34 & -1.86\\
2.94 &  0.58 & 0.378 & 5.64 & -1.80\\
3.41 & -0.10 & 0.360 & 5.04 & -1.45\\
4.03 & -0.99 & 0.338 & 4.37 & -0.63\\
\end{tabular}
\end{ruledtabular}
\end{table}

Next, we demonstrate the accuracy of our approach for kentropies. For our example, 
\Eq{Kscc} simplifies to 
\ben\label{Kscc.model}
\breve{K}_{{\sss S},\breve{n}\s^{\sss \T}} ^{\sss \T,{\rm cc}}[v] =
K_{\sss S ,0}^{\sss \T} 
+\int dx~\left\{\breve{\bar{n}}\s^{\sss \T}(x)-\breve{n}\s^{\sss \T}[v](x)\right\}v(x)\ .
\een
In this case the reference potential is not zero, but an infinite square well.
Hence, a kentropic contribution 
$K_{\sss S ,0}^{\sss \T} = T_{\sss S ,0}^{\sss \T} - S_{\sss S ,0}^{\sss \T}/\beta$ 
of the reference system appears, which we can compute exactly quite simply. 
The kinetic energy of the infinite square well is 
$T^{\sss \T}_{\sss S ,0}=\sum_j^N f_j^{\sss \T}\,\e_{j,0}$, 
and the entropy is 
$S^{\sss \T}_{\sss S ,0}=- \sum_j {f_j^{\sss \T} \ln(f_j^{\sss \T}) 
+ (1-f_j^{\sss \T}) \ln(1-f_j^{\sss \T})}$,
with $f_j^{\sss \T} = 1/(1.0+\exp{[\beta(\e_{j,0}-\mu_0)]})$ denoting Fermi functions
and $\e_{j,0}$ and $\mu_0$ the $j^{\rm th}$ eigenvalue and chemical potential 
for the non-interacting reference system. We avoid temperature-dependent KS eigenvalues\cite{PPGB14} 
by choosing a purely non-interacting reference system, not a KS system associated with 
a specific interacting system. 
Evaluating \Eq{Kscc.model} for the same potential as in Fig.~\ref{f:n}
yields the results in Tab.~\ref{t:K.L10D2m1}.
We measure the error of TF theory and our PFA with respect to the residual kentropy 
$\Delta K\s^{\sss \T} = K_{{\sss S},n\s^{\sss \T}}^{\sss \T, cc}-K_{\sss S ,0}^{\sss \T}$,
because this is the only piece of the total kentropy being approximated. 
We also list the electron degeneracy parameter values 
and the kentropic contributions of the reference system.
Reaching from cold temperatures up to the WDM regime
(around $\Lambda\approx 1$) our PFA yields kentropies that are significantly
more accurate than TF theory, improving them by roughly an order of magnitude. 
Once we go far beyond the WDM regime where the entropic contribution starts dominating, 
the errors would become comparable. 

\begin{figure}[htb]%
\begin{center}
\includegraphics[width=8.5cm]{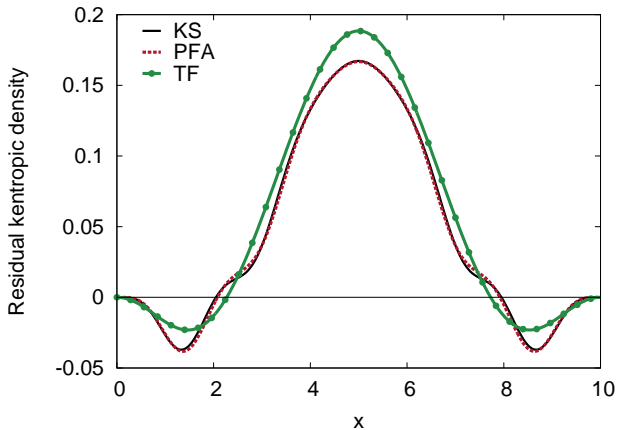}
\end{center}
\caption {Residual kentropic density of five particles 
in the same potential as in Fig.~\ref{f:n} in the WDM regime.
Our PFA (solid red curve) derived in Eq.~(\ref{n.finT})
is on top of the exact result (solid black curve). 
The TF result (dotted green curve), on the other hand, 
follows the general trend as expected, but misses quantitative details.}
\label{f:Kscc.int.N5.temp03}%
\end{figure}

We can better understand the advantage of our PFA  over the conventional TF approach 
by analyzing both in real space. We compute residual kentropic densities
(the integrand of \Eq{Kscc.model}) for the example in Fig.~\ref{f:n}. 
As illustrated in Fig.~\ref{f:Kscc.int.N5.temp03}. The TF approach (dotted green curve) 
only reproduces the qualitative trends of the exact result (black curve). Nevertheless, 
TF theory gives reasonable results for the kentropy (see Tab.~\ref{t:K.L10D2m1}),
but only due to an apparent error cancellation. 
As shown in the figure, the error due to an overestimation in the interior is balanced 
by an underestimation in the outer regions of the system. 
Our PFA, on the other hand, not only yields accurate integrated kentropies (area under the curve 
in Fig.~\ref{f:Kscc.int.N5.temp03}), but is also highly accurate in real space.  As such, 
and unlike TF, our PFA does not rely on cancellation of errors in the kentropy density for its accurate kentropy values.

The finite-temperature PFA approach outlined here offers several
advantages over other methods, particularly for WDM, where solution of the KS
equations for numerous occupied states becomes especially daunting.  We retain
the advantages of the KS system while avoiding the costly, repetitive solution
of eigenvalue problems by isolating a small piece of the kentropy to
approximate through the coupling-constant formalism.  The reference system is
always chosen such that its kentropy is known exactly.  Combined with our
density approximation, this improves approximate kentropies by up to an order
of magnitude in the WDM regime and produces highly accurate kentropic
densities.  The density approximation derived in this paper is computationally
efficient because only the leading terms are needed for convergence at
WDM temperatures.

The path integral method used to derive this approximation\cite{CSB14} invites
use of successful zero-temperature approximations to the propagator, and it is
a promising approach for extension to three dimensional systems.  
Furthermore, combining the presented finite-temperature PFT with semiclassical methods 
offers prospects for a systematic route to exchange
energy approximations, instead of only relying on existing, 
zero-temperature density functional approximations.  Work in this direction is currently in development.
With these advantages, finite-temperature PFT is poised to bridge the
``malfunction junction" of WDM by providing computationally efficient,
semiclassical methods at high temperatures and densities.

\section*{Acknowledgments}
We acknowledge Hardy Gross and Kieron Burke for 
providing a fruitful atmosphere facilitating independent research. 
We are grateful to Rudy Magyar for useful discussion. 
A.C. has been partially supported by NSF grant CHE-1112442.  
A.P.J. is supported by DOE grant DE-FG02-97ER25308. 

\bibliography{ftpft}

\begin{thebibliography}{39}%
\makeatletter
\providecommand \@ifxundefined [1]{%
 \@ifx{#1\undefined}
}%
\providecommand \@ifnum [1]{%
 \ifnum #1\expandafter \@firstoftwo
 \else \expandafter \@secondoftwo
 \fi
}%
\providecommand \@ifx [1]{%
 \ifx #1\expandafter \@firstoftwo
 \else \expandafter \@secondoftwo
 \fi
}%
\providecommand \natexlab [1]{#1}%
\providecommand \enquote  [1]{``#1''}%
\providecommand \bibnamefont  [1]{#1}%
\providecommand \bibfnamefont [1]{#1}%
\providecommand \citenamefont [1]{#1}%
\providecommand \href@noop [0]{\@secondoftwo}%
\providecommand \href [0]{\begingroup \@sanitize@url \@href}%
\providecommand \@href[1]{\@@startlink{#1}\@@href}%
\providecommand \@@href[1]{\endgroup#1\@@endlink}%
\providecommand \@sanitize@url [0]{\catcode `\\12\catcode `\$12\catcode
  `\&12\catcode `\#12\catcode `\^12\catcode `\_12\catcode `\%12\relax}%
\providecommand \@@startlink[1]{}%
\providecommand \@@endlink[0]{}%
\providecommand \url  [0]{\begingroup\@sanitize@url \@url }%
\providecommand \@url [1]{\endgroup\@href {#1}{\urlprefix }}%
\providecommand \urlprefix  [0]{URL }%
\providecommand \Eprint [0]{\href }%
\providecommand \doibase [0]{http://dx.doi.org/}%
\providecommand \selectlanguage [0]{\@gobble}%
\providecommand \bibinfo  [0]{\@secondoftwo}%
\providecommand \bibfield  [0]{\@secondoftwo}%
\providecommand \translation [1]{[#1]}%
\providecommand \BibitemOpen [0]{}%
\providecommand \bibitemStop [0]{}%
\providecommand \bibitemNoStop [0]{.\EOS\space}%
\providecommand \EOS [0]{\spacefactor3000\relax}%
\providecommand \BibitemShut  [1]{\csname bibitem#1\endcsname}%
\let\auto@bib@innerbib\@empty
\bibitem [{\citenamefont {Graziani}\ \emph {et~al.}(2014)\citenamefont
  {Graziani}, \citenamefont {Desjarlais}, \citenamefont {Redmer},\ and\
  \citenamefont {Trickey}}]{GDRT14}%
  \BibitemOpen
  \bibinfo {editor} {\bibfnamefont {F.}~\bibnamefont {Graziani}}, \bibinfo
  {editor} {\bibfnamefont {M.~P.}\ \bibnamefont {Desjarlais}}, \bibinfo
  {editor} {\bibfnamefont {R.}~\bibnamefont {Redmer}}, \ and\ \bibinfo {editor}
  {\bibfnamefont {S.~B.}\ \bibnamefont {Trickey}},\ eds.,\ \href@noop {} {\emph
  {\bibinfo {title} {Frontiers and Challenges in Warm Dense Matter}}},\
  \bibinfo {series} {Lecture Notes in Computational Science and Engineering},
  Vol.~\bibinfo {volume} {96}\ (\bibinfo  {publisher} {Springer International
  Publishing},\ \bibinfo {year} {2014})\BibitemShut {NoStop}%
\bibitem [{\citenamefont {Atzeni}\ and\ \citenamefont {Meyer-ter
  Vehn}(2004)}]{A04}%
  \BibitemOpen
  \bibfield  {author} {\bibinfo {author} {\bibfnamefont {S.}~\bibnamefont
  {Atzeni}}\ and\ \bibinfo {author} {\bibfnamefont {J.}~\bibnamefont {Meyer-ter
  Vehn}},\ }\href@noop {} {\emph {\bibinfo {title} {The Physics of Inertial
  Fusion: Beam-Plasma Interaction, Hydrodynamics, Hot Dense Matter}}}\
  (\bibinfo  {publisher} {Clarendon Press},\ \bibinfo {year}
  {2004})\BibitemShut {NoStop}%
\bibitem [{\citenamefont {on~High Energy Density Plasma Physics Plasma
  Science~Committee}(2003)}]{HEDP03}%
  \BibitemOpen
  \bibfield  {author} {\bibinfo {author} {\bibfnamefont {N.~R. C.~C.}\
  \bibnamefont {on~High Energy Density Plasma Physics Plasma
  Science~Committee}},\ }\href@noop {} {\emph {\bibinfo {title} {Frontiers in
  High Energy Density Physics: The X-Games of Contemporary Science}}}\
  (\bibinfo  {publisher} {The National Academies Press},\ \bibinfo {year}
  {2003})\BibitemShut {NoStop}%
\bibitem [{\citenamefont {Mattsson}\ and\ \citenamefont
  {Desjarlais}(2006)}]{MD06}%
  \BibitemOpen
  \bibfield  {author} {\bibinfo {author} {\bibfnamefont {T.~R.}\ \bibnamefont
  {Mattsson}}\ and\ \bibinfo {author} {\bibfnamefont {M.~P.}\ \bibnamefont
  {Desjarlais}},\ }\href@noop {} {\bibfield  {journal} {\bibinfo  {journal}
  {Phys. Rev. Lett.}\ }\textbf {\bibinfo {volume} {97}},\ \bibinfo {pages}
  {017801} (\bibinfo {year} {2006})}\BibitemShut {NoStop}%
\bibitem [{\citenamefont {Graziani}\ \emph {et~al.}(2012)\citenamefont
  {Graziani}, \citenamefont {Batista}, \citenamefont {Benedict}, \citenamefont
  {Castor}, \citenamefont {Chen}, \citenamefont {Chen}, \citenamefont {Fichtl},
  \citenamefont {Glosli}, \citenamefont {Grabowski}, \citenamefont {Graf},
  \citenamefont {Hau-Riege}, \citenamefont {Hazi}, \citenamefont {Khairallah},
  \citenamefont {Krauss}, \citenamefont {Langdon}, \citenamefont {London},
  \citenamefont {Markmann}, \citenamefont {Murillo}, \citenamefont {Richards},
  \citenamefont {Scott}, \citenamefont {Shepherd}, \citenamefont {Stanton},
  \citenamefont {Streitz}, \citenamefont {Surh}, \citenamefont {Weisheit},\
  and\ \citenamefont {Whitley}}]{GBBC12}%
  \BibitemOpen
  \bibfield  {author} {\bibinfo {author} {\bibfnamefont {F.~R.}\ \bibnamefont
  {Graziani}}, \bibinfo {author} {\bibfnamefont {V.~S.}\ \bibnamefont
  {Batista}}, \bibinfo {author} {\bibfnamefont {L.~X.}\ \bibnamefont
  {Benedict}}, \bibinfo {author} {\bibfnamefont {J.~I.}\ \bibnamefont
  {Castor}}, \bibinfo {author} {\bibfnamefont {H.}~\bibnamefont {Chen}},
  \bibinfo {author} {\bibfnamefont {S.~N.}\ \bibnamefont {Chen}}, \bibinfo
  {author} {\bibfnamefont {C.~A.}\ \bibnamefont {Fichtl}}, \bibinfo {author}
  {\bibfnamefont {J.~N.}\ \bibnamefont {Glosli}}, \bibinfo {author}
  {\bibfnamefont {P.~E.}\ \bibnamefont {Grabowski}}, \bibinfo {author}
  {\bibfnamefont {A.~T.}\ \bibnamefont {Graf}}, \bibinfo {author}
  {\bibfnamefont {S.~P.}\ \bibnamefont {Hau-Riege}}, \bibinfo {author}
  {\bibfnamefont {A.~U.}\ \bibnamefont {Hazi}}, \bibinfo {author}
  {\bibfnamefont {S.~A.}\ \bibnamefont {Khairallah}}, \bibinfo {author}
  {\bibfnamefont {L.}~\bibnamefont {Krauss}}, \bibinfo {author} {\bibfnamefont
  {A.~B.}\ \bibnamefont {Langdon}}, \bibinfo {author} {\bibfnamefont {R.~A.}\
  \bibnamefont {London}}, \bibinfo {author} {\bibfnamefont {A.}~\bibnamefont
  {Markmann}}, \bibinfo {author} {\bibfnamefont {M.~S.}\ \bibnamefont
  {Murillo}}, \bibinfo {author} {\bibfnamefont {D.~F.}\ \bibnamefont
  {Richards}}, \bibinfo {author} {\bibfnamefont {H.~A.}\ \bibnamefont {Scott}},
  \bibinfo {author} {\bibfnamefont {R.}~\bibnamefont {Shepherd}}, \bibinfo
  {author} {\bibfnamefont {L.~G.}\ \bibnamefont {Stanton}}, \bibinfo {author}
  {\bibfnamefont {F.~H.}\ \bibnamefont {Streitz}}, \bibinfo {author}
  {\bibfnamefont {M.~P.}\ \bibnamefont {Surh}}, \bibinfo {author}
  {\bibfnamefont {J.~C.}\ \bibnamefont {Weisheit}}, \ and\ \bibinfo {author}
  {\bibfnamefont {H.~D.}\ \bibnamefont {Whitley}},\ }\href@noop {} {\bibfield
  {journal} {\bibinfo  {journal} {High Energy Density Physics}\ }\textbf
  {\bibinfo {volume} {8}},\ \bibinfo {pages} {105 } (\bibinfo {year}
  {2012})}\BibitemShut {NoStop}%
\bibitem [{\citenamefont {Sanbonmatsu}\ \emph {et~al.}(2000)\citenamefont
  {Sanbonmatsu}, \citenamefont {Thode}, \citenamefont {Vu},\ and\ \citenamefont
  {Murillo}}]{STVM00}%
  \BibitemOpen
  \bibfield  {author} {\bibinfo {author} {\bibfnamefont {K.~Y.}\ \bibnamefont
  {Sanbonmatsu}}, \bibinfo {author} {\bibfnamefont {L.~E.}\ \bibnamefont
  {Thode}}, \bibinfo {author} {\bibfnamefont {H.~X.}\ \bibnamefont {Vu}}, \
  and\ \bibinfo {author} {\bibfnamefont {M.~S.}\ \bibnamefont {Murillo}},\
  }\href@noop {} {\bibfield  {journal} {\bibinfo  {journal} {J. Phys. IV
  France}\ }\textbf {\bibinfo {volume} {10}},\ \bibinfo {pages} {Pr5} (\bibinfo
  {year} {2000})}\BibitemShut {NoStop}%
\bibitem [{\citenamefont {Knudson}\ and\ \citenamefont
  {Desjarlais}(2009)}]{KD09}%
  \BibitemOpen
  \bibfield  {author} {\bibinfo {author} {\bibfnamefont {M.~D.}\ \bibnamefont
  {Knudson}}\ and\ \bibinfo {author} {\bibfnamefont {M.~P.}\ \bibnamefont
  {Desjarlais}},\ }\href@noop {} {\bibfield  {journal} {\bibinfo  {journal}
  {Phys. Rev. Lett.}\ }\textbf {\bibinfo {volume} {103}},\ \bibinfo {pages}
  {225501} (\bibinfo {year} {2009})}\BibitemShut {NoStop}%
\bibitem [{\citenamefont {Holst}\ \emph {et~al.}(2008)\citenamefont {Holst},
  \citenamefont {Redmer},\ and\ \citenamefont {Desjarlais}}]{HRD08}%
  \BibitemOpen
  \bibfield  {author} {\bibinfo {author} {\bibfnamefont {B.}~\bibnamefont
  {Holst}}, \bibinfo {author} {\bibfnamefont {R.}~\bibnamefont {Redmer}}, \
  and\ \bibinfo {author} {\bibfnamefont {M.~P.}\ \bibnamefont {Desjarlais}},\
  }\href@noop {} {\bibfield  {journal} {\bibinfo  {journal} {Phys. Rev. B}\
  }\textbf {\bibinfo {volume} {77}},\ \bibinfo {pages} {184201} (\bibinfo
  {year} {2008})}\BibitemShut {NoStop}%
\bibitem [{\citenamefont {Hohenberg}\ and\ \citenamefont {Kohn}(1964)}]{HK64}%
  \BibitemOpen
  \bibfield  {author} {\bibinfo {author} {\bibfnamefont {P.}~\bibnamefont
  {Hohenberg}}\ and\ \bibinfo {author} {\bibfnamefont {W.}~\bibnamefont
  {Kohn}},\ }\href {\doibase 10.1103/PhysRev.136.B864} {\bibfield  {journal}
  {\bibinfo  {journal} {Phys. Rev.}\ }\textbf {\bibinfo {volume} {136}},\
  \bibinfo {pages} {B864} (\bibinfo {year} {1964})}\BibitemShut {NoStop}%
\bibitem [{\citenamefont {Kohn}\ and\ \citenamefont {Sham}(1965)}]{KS65}%
  \BibitemOpen
  \bibfield  {author} {\bibinfo {author} {\bibfnamefont {W.}~\bibnamefont
  {Kohn}}\ and\ \bibinfo {author} {\bibfnamefont {L.~J.}\ \bibnamefont
  {Sham}},\ }\href {\doibase 10.1103/PhysRev.140.A1133} {\bibfield  {journal}
  {\bibinfo  {journal} {Phys. Rev.}\ }\textbf {\bibinfo {volume} {140}},\
  \bibinfo {pages} {A1133} (\bibinfo {year} {1965})}\BibitemShut {NoStop}%
\bibitem [{\citenamefont {Kietzmann}\ \emph {et~al.}(2008)\citenamefont
  {Kietzmann}, \citenamefont {Redmer}, \citenamefont {Desjarlais},\ and\
  \citenamefont {Mattsson}}]{KRDM08}%
  \BibitemOpen
  \bibfield  {author} {\bibinfo {author} {\bibfnamefont {A.}~\bibnamefont
  {Kietzmann}}, \bibinfo {author} {\bibfnamefont {R.}~\bibnamefont {Redmer}},
  \bibinfo {author} {\bibfnamefont {M.~P.}\ \bibnamefont {Desjarlais}}, \ and\
  \bibinfo {author} {\bibfnamefont {T.~R.}\ \bibnamefont {Mattsson}},\
  }\href@noop {} {\bibfield  {journal} {\bibinfo  {journal} {Phys. Rev. Lett.}\
  }\textbf {\bibinfo {volume} {101}},\ \bibinfo {pages} {070401} (\bibinfo
  {year} {2008})}\BibitemShut {NoStop}%
\bibitem [{\citenamefont {Root}\ \emph {et~al.}(2010)\citenamefont {Root},
  \citenamefont {Magyar}, \citenamefont {Carpenter}, \citenamefont {Hanson},\
  and\ \citenamefont {Mattsson}}]{RMCH10}%
  \BibitemOpen
  \bibfield  {author} {\bibinfo {author} {\bibfnamefont {S.}~\bibnamefont
  {Root}}, \bibinfo {author} {\bibfnamefont {R.~J.}\ \bibnamefont {Magyar}},
  \bibinfo {author} {\bibfnamefont {J.~H.}\ \bibnamefont {Carpenter}}, \bibinfo
  {author} {\bibfnamefont {D.~L.}\ \bibnamefont {Hanson}}, \ and\ \bibinfo
  {author} {\bibfnamefont {T.~R.}\ \bibnamefont {Mattsson}},\ }\href@noop {}
  {\bibfield  {journal} {\bibinfo  {journal} {Phys. Rev. Lett.}\ }\textbf
  {\bibinfo {volume} {105}},\ \bibinfo {pages} {085501} (\bibinfo {year}
  {2010})}\BibitemShut {NoStop}%
\bibitem [{\citenamefont {Marx}\ and\ \citenamefont {Hutter}(2000)}]{MH00}%
  \BibitemOpen
  \bibfield  {author} {\bibinfo {author} {\bibfnamefont {D.}~\bibnamefont
  {Marx}}\ and\ \bibinfo {author} {\bibfnamefont {J.}~\bibnamefont {Hutter}},\
  }in\ \href@noop {} {\emph {\bibinfo {booktitle} {Modern Methods and
  Algorithms of Quantum Chemistry}}},\ \bibinfo {series} {NIC Series},
  Vol.~\bibinfo {volume} {1},\ \bibinfo {editor} {edited by\ \bibinfo {editor}
  {\bibfnamefont {J.}~\bibnamefont {Grotendorst}}}\ (\bibinfo  {publisher}
  {J\:{u}lich},\ \bibinfo {year} {2000})\ pp.\ \bibinfo {pages}
  {301--449}\BibitemShut {NoStop}%
\bibitem [{\citenamefont {Marx}\ and\ \citenamefont {Hutter}(2009)}]{MH09}%
  \BibitemOpen
  \bibfield  {author} {\bibinfo {author} {\bibfnamefont {D.}~\bibnamefont
  {Marx}}\ and\ \bibinfo {author} {\bibfnamefont {J.}~\bibnamefont {Hutter}},\
  }\href@noop {} {\emph {\bibinfo {title} {Ab Initio Molecular Dynamics: Basic
  Theory and Advanced Methods}}}\ (\bibinfo  {publisher} {Cambridge University
  Press},\ \bibinfo {year} {2009})\BibitemShut {NoStop}%
\bibitem [{\citenamefont {Karasiev}\ \emph {et~al.}(2014)\citenamefont
  {Karasiev}, \citenamefont {Sjostrom}, \citenamefont {Chakraborty},
  \citenamefont {Dufty}, \citenamefont {Runge}, \citenamefont {Harris},\ and\
  \citenamefont {Trickey}}]{KSCD14}%
  \BibitemOpen
  \bibfield  {author} {\bibinfo {author} {\bibfnamefont {V.~V.}\ \bibnamefont
  {Karasiev}}, \bibinfo {author} {\bibfnamefont {T.}~\bibnamefont {Sjostrom}},
  \bibinfo {author} {\bibfnamefont {D.}~\bibnamefont {Chakraborty}}, \bibinfo
  {author} {\bibfnamefont {J.~W.}\ \bibnamefont {Dufty}}, \bibinfo {author}
  {\bibfnamefont {K.}~\bibnamefont {Runge}}, \bibinfo {author} {\bibfnamefont
  {F.~E.}\ \bibnamefont {Harris}}, \ and\ \bibinfo {author} {\bibfnamefont
  {S.}~\bibnamefont {Trickey}},\ }in\ \href@noop {} {\emph {\bibinfo
  {booktitle} {Frontiers and Challenges in Warm Dense Matter}}},\ \bibinfo
  {series} {Lecture Notes in Computational Science and Engineering},
  Vol.~\bibinfo {volume} {96},\ \bibinfo {editor} {edited by\ \bibinfo {editor}
  {\bibfnamefont {F.}~\bibnamefont {Graziani}}, \bibinfo {editor}
  {\bibfnamefont {M.~P.}\ \bibnamefont {Desjarlais}}, \bibinfo {editor}
  {\bibfnamefont {R.}~\bibnamefont {Redmer}}, \ and\ \bibinfo {editor}
  {\bibfnamefont {S.~B.}\ \bibnamefont {Trickey}}}\ (\bibinfo  {publisher}
  {Springer International Publishing},\ \bibinfo {year} {2014})\ pp.\ \bibinfo
  {pages} {61--85}\BibitemShut {NoStop}%
\bibitem [{\citenamefont {Wang}\ and\ \citenamefont {Carter}(2000)}]{WC00}%
  \BibitemOpen
  \bibfield  {author} {\bibinfo {author} {\bibfnamefont {Y.~A.}\ \bibnamefont
  {Wang}}\ and\ \bibinfo {author} {\bibfnamefont {E.~A.}\ \bibnamefont
  {Carter}},\ }in\ \href@noop {} {\emph {\bibinfo {booktitle} {Theoretical
  Methods in Condensed Phase Chemistry}}},\ \bibinfo {editor} {edited by\
  \bibinfo {editor} {\bibfnamefont {S.~D.}\ \bibnamefont {Schwartz}}}\
  (\bibinfo  {publisher} {Kluwer},\ \bibinfo {address} {Dordrecht},\ \bibinfo
  {year} {2000})\ Chap.~\bibinfo {chapter} {5}, p.\ \bibinfo {pages}
  {117}\BibitemShut {NoStop}%
\bibitem [{\citenamefont {Karasiev}\ \emph {et~al.}(2012)\citenamefont
  {Karasiev}, \citenamefont {Sjostrom},\ and\ \citenamefont {Trickey}}]{KST12}%
  \BibitemOpen
  \bibfield  {author} {\bibinfo {author} {\bibfnamefont {V.~V.}\ \bibnamefont
  {Karasiev}}, \bibinfo {author} {\bibfnamefont {T.}~\bibnamefont {Sjostrom}},
  \ and\ \bibinfo {author} {\bibfnamefont {S.~B.}\ \bibnamefont {Trickey}},\
  }\href@noop {} {\bibfield  {journal} {\bibinfo  {journal} {Phys. Rev. B}\
  }\textbf {\bibinfo {volume} {86}},\ \bibinfo {pages} {115101} (\bibinfo
  {year} {2012})}\BibitemShut {NoStop}%
\bibitem [{\citenamefont {Karasiev}\ \emph {et~al.}(2013)\citenamefont
  {Karasiev}, \citenamefont {Chakraborty}, \citenamefont {Shukruto},\ and\
  \citenamefont {Trickey}}]{KCST13}%
  \BibitemOpen
  \bibfield  {author} {\bibinfo {author} {\bibfnamefont {V.~V.}\ \bibnamefont
  {Karasiev}}, \bibinfo {author} {\bibfnamefont {D.}~\bibnamefont
  {Chakraborty}}, \bibinfo {author} {\bibfnamefont {O.~A.}\ \bibnamefont
  {Shukruto}}, \ and\ \bibinfo {author} {\bibfnamefont {S.~B.}\ \bibnamefont
  {Trickey}},\ }\href@noop {} {\bibfield  {journal} {\bibinfo  {journal} {Phys.
  Rev. B}\ }\textbf {\bibinfo {volume} {88}},\ \bibinfo {pages} {161108}
  (\bibinfo {year} {2013})}\BibitemShut {NoStop}%
\bibitem [{\citenamefont {Sjostrom}\ and\ \citenamefont
  {Daligault}(2013)}]{SD13}%
  \BibitemOpen
  \bibfield  {author} {\bibinfo {author} {\bibfnamefont {T.}~\bibnamefont
  {Sjostrom}}\ and\ \bibinfo {author} {\bibfnamefont {J.}~\bibnamefont
  {Daligault}},\ }\href@noop {} {\bibfield  {journal} {\bibinfo  {journal}
  {Phys. Rev. B}\ }\textbf {\bibinfo {volume} {88}},\ \bibinfo {pages} {195103}
  (\bibinfo {year} {2013})}\BibitemShut {NoStop}%
\bibitem [{\citenamefont {Cangi}\ \emph {et~al.}(2011)\citenamefont {Cangi},
  \citenamefont {Lee}, \citenamefont {Elliott}, \citenamefont {Burke},\ and\
  \citenamefont {Gross}}]{CLEB11}%
  \BibitemOpen
  \bibfield  {author} {\bibinfo {author} {\bibfnamefont {A.}~\bibnamefont
  {Cangi}}, \bibinfo {author} {\bibfnamefont {D.}~\bibnamefont {Lee}}, \bibinfo
  {author} {\bibfnamefont {P.}~\bibnamefont {Elliott}}, \bibinfo {author}
  {\bibfnamefont {K.}~\bibnamefont {Burke}}, \ and\ \bibinfo {author}
  {\bibfnamefont {E.~K.~U.}\ \bibnamefont {Gross}},\ }\href {\doibase
  10.1103/PhysRevLett.106.236404} {\bibfield  {journal} {\bibinfo  {journal}
  {Phys. Rev. Lett.}\ }\textbf {\bibinfo {volume} {106}},\ \bibinfo {pages}
  {236404} (\bibinfo {year} {2011})}\BibitemShut {NoStop}%
\bibitem [{\citenamefont {Cangi}\ \emph {et~al.}(2013)\citenamefont {Cangi},
  \citenamefont {Gross},\ and\ \citenamefont {Burke}}]{CGB13}%
  \BibitemOpen
  \bibfield  {author} {\bibinfo {author} {\bibfnamefont {A.}~\bibnamefont
  {Cangi}}, \bibinfo {author} {\bibfnamefont {E.~K.~U.}\ \bibnamefont {Gross}},
  \ and\ \bibinfo {author} {\bibfnamefont {K.}~\bibnamefont {Burke}},\ }\href
  {\doibase 10.1103/PhysRevA.88.062505} {\bibfield  {journal} {\bibinfo
  {journal} {Phys. Rev. A}\ }\textbf {\bibinfo {volume} {88}},\ \bibinfo
  {pages} {062505} (\bibinfo {year} {2013})}\BibitemShut {NoStop}%
\bibitem [{\citenamefont {Cangi}\ \emph {et~al.}(2010)\citenamefont {Cangi},
  \citenamefont {Lee}, \citenamefont {Elliott},\ and\ \citenamefont
  {Burke}}]{CLEB10}%
  \BibitemOpen
  \bibfield  {author} {\bibinfo {author} {\bibfnamefont {A.}~\bibnamefont
  {Cangi}}, \bibinfo {author} {\bibfnamefont {D.}~\bibnamefont {Lee}}, \bibinfo
  {author} {\bibfnamefont {P.}~\bibnamefont {Elliott}}, \ and\ \bibinfo
  {author} {\bibfnamefont {K.}~\bibnamefont {Burke}},\ }\href {\doibase
  10.1103/PhysRevB.81.235128} {\bibfield  {journal} {\bibinfo  {journal} {Phys.
  Rev. B}\ }\textbf {\bibinfo {volume} {81}},\ \bibinfo {pages} {235128}
  (\bibinfo {year} {2010})}\BibitemShut {NoStop}%
\bibitem [{\citenamefont {Elliott}\ \emph {et~al.}(2008)\citenamefont
  {Elliott}, \citenamefont {Lee}, \citenamefont {Cangi},\ and\ \citenamefont
  {Burke}}]{ELCB08}%
  \BibitemOpen
  \bibfield  {author} {\bibinfo {author} {\bibfnamefont {P.}~\bibnamefont
  {Elliott}}, \bibinfo {author} {\bibfnamefont {D.}~\bibnamefont {Lee}},
  \bibinfo {author} {\bibfnamefont {A.}~\bibnamefont {Cangi}}, \ and\ \bibinfo
  {author} {\bibfnamefont {K.}~\bibnamefont {Burke}},\ }\href {\doibase
  10.1103/PhysRevLett.100.256406} {\bibfield  {journal} {\bibinfo  {journal}
  {Phys. Rev. Lett.}\ }\textbf {\bibinfo {volume} {100}},\ \bibinfo {pages}
  {256406} (\bibinfo {year} {2008})}\BibitemShut {NoStop}%
\bibitem [{\citenamefont {Roccia}\ \emph {et~al.}(2010)\citenamefont {Roccia},
  \citenamefont {Brack},\ and\ \citenamefont {Koch}}]{RBK10}%
  \BibitemOpen
  \bibfield  {author} {\bibinfo {author} {\bibfnamefont {J.}~\bibnamefont
  {Roccia}}, \bibinfo {author} {\bibfnamefont {M.}~\bibnamefont {Brack}}, \
  and\ \bibinfo {author} {\bibfnamefont {A.}~\bibnamefont {Koch}},\ }\href
  {\doibase 10.1103/PhysRevE.81.011118} {\bibfield  {journal} {\bibinfo
  {journal} {Phys. Rev. E}\ }\textbf {\bibinfo {volume} {81}},\ \bibinfo
  {pages} {011118} (\bibinfo {year} {2010})}\BibitemShut {NoStop}%
\bibitem [{\citenamefont {Baer}\ \emph {et~al.}(2013)\citenamefont {Baer},
  \citenamefont {Neuhauser},\ and\ \citenamefont {Rabani}}]{BNR13}%
  \BibitemOpen
  \bibfield  {author} {\bibinfo {author} {\bibfnamefont {R.}~\bibnamefont
  {Baer}}, \bibinfo {author} {\bibfnamefont {D.}~\bibnamefont {Neuhauser}}, \
  and\ \bibinfo {author} {\bibfnamefont {E.}~\bibnamefont {Rabani}},\ }\href
  {\doibase 10.1103/PhysRevLett.111.106402} {\bibfield  {journal} {\bibinfo
  {journal} {Phys. Rev. Lett.}\ }\textbf {\bibinfo {volume} {111}},\ \bibinfo
  {pages} {106402} (\bibinfo {year} {2013})}\BibitemShut {NoStop}%
\bibitem [{\citenamefont {Lieb}\ and\ \citenamefont {Simon}(1973)}]{LS73}%
  \BibitemOpen
  \bibfield  {author} {\bibinfo {author} {\bibfnamefont {E.}~\bibnamefont
  {Lieb}}\ and\ \bibinfo {author} {\bibfnamefont {B.}~\bibnamefont {Simon}},\
  }\href@noop {} {\bibfield  {journal} {\bibinfo  {journal} {Phys. Rev. Lett.}\
  }\textbf {\bibinfo {volume} {31}},\ \bibinfo {pages} {681} (\bibinfo {year}
  {1973})}\BibitemShut {NoStop}%
\bibitem [{\citenamefont {Thomas}(1927)}]{T27}%
  \BibitemOpen
  \bibfield  {author} {\bibinfo {author} {\bibfnamefont {L.}~\bibnamefont
  {Thomas}},\ }\href {\doibase 10.1017/S0305004100011683} {\bibfield  {journal}
  {\bibinfo  {journal} {Math. Proc. Camb. Phil. Soc.}\ }\textbf {\bibinfo
  {volume} {23}},\ \bibinfo {pages} {542} (\bibinfo {year} {1927})}\BibitemShut
  {NoStop}%
\bibitem [{\citenamefont {Fermi}(1928)}]{F28}%
  \BibitemOpen
  \bibfield  {author} {\bibinfo {author} {\bibfnamefont {E.}~\bibnamefont
  {Fermi}},\ }\href {http://dx.doi.org/10.1007/BF01351576} {\bibfield
  {journal} {\bibinfo  {journal} {Z. Phys. A}\ }\textbf {\bibinfo {volume}
  {48}},\ \bibinfo {pages} {73} (\bibinfo {year} {1928})}\BibitemShut {NoStop}%
\bibitem [{\citenamefont {Narnhofer}\ and\ \citenamefont
  {Thirring}(1981)}]{NT80}%
  \BibitemOpen
  \bibfield  {author} {\bibinfo {author} {\bibfnamefont {H.}~\bibnamefont
  {Narnhofer}}\ and\ \bibinfo {author} {\bibfnamefont {W.}~\bibnamefont
  {Thirring}},\ }\href {\doibase
  http://dx.doi.org/10.1016/0003-4916(81)90008-7} {\bibfield  {journal}
  {\bibinfo  {journal} {Annals of Physics}\ }\textbf {\bibinfo {volume}
  {134}},\ \bibinfo {pages} {128 } (\bibinfo {year} {1981})}\BibitemShut
  {NoStop}%
\bibitem [{\citenamefont {Pribram-Jones}\ \emph {et~al.}(2014)\citenamefont
  {Pribram-Jones}, \citenamefont {Pittalis}, \citenamefont {Gross},\ and\
  \citenamefont {Burke}}]{PPGB14}%
  \BibitemOpen
  \bibfield  {author} {\bibinfo {author} {\bibfnamefont {A.}~\bibnamefont
  {Pribram-Jones}}, \bibinfo {author} {\bibfnamefont {S.}~\bibnamefont
  {Pittalis}}, \bibinfo {author} {\bibfnamefont {E.}~\bibnamefont {Gross}}, \
  and\ \bibinfo {author} {\bibfnamefont {K.}~\bibnamefont {Burke}},\ }in\ \href
  {\doibase 10.1007/978-3-319-04912-0_2} {\emph {\bibinfo {booktitle}
  {Frontiers and Challenges in Warm Dense Matter}}},\ \bibinfo {series}
  {Lecture Notes in Computational Science and Engineering}, Vol.~\bibinfo
  {volume} {96},\ \bibinfo {editor} {edited by\ \bibinfo {editor}
  {\bibfnamefont {F.}~\bibnamefont {Graziani}}, \bibinfo {editor}
  {\bibfnamefont {M.~P.}\ \bibnamefont {Desjarlais}}, \bibinfo {editor}
  {\bibfnamefont {R.}~\bibnamefont {Redmer}}, \ and\ \bibinfo {editor}
  {\bibfnamefont {S.~B.}\ \bibnamefont {Trickey}}}\ (\bibinfo  {publisher}
  {Springer International Publishing},\ \bibinfo {year} {2014})\ pp.\ \bibinfo
  {pages} {25--60}\BibitemShut {NoStop}%
\bibitem [{\citenamefont {Mermin}(1965)}]{M65}%
  \BibitemOpen
  \bibfield  {author} {\bibinfo {author} {\bibfnamefont {N.~D.}\ \bibnamefont
  {Mermin}},\ }\href@noop {} {\bibfield  {journal} {\bibinfo  {journal} {Phys.
  Rev.}\ }\textbf {\bibinfo {volume} {137}},\ \bibinfo {pages} {A: 1441}
  (\bibinfo {year} {1965})}\BibitemShut {NoStop}%
\bibitem [{\citenamefont {Eschrig}(2010)}]{E10}%
  \BibitemOpen
  \bibfield  {author} {\bibinfo {author} {\bibfnamefont {H.}~\bibnamefont
  {Eschrig}},\ }\href@noop {} {\bibfield  {journal} {\bibinfo  {journal} {Phys.
  Rev. B}\ }\textbf {\bibinfo {volume} {82}},\ \bibinfo {pages} {205120}
  (\bibinfo {year} {2010})}\BibitemShut {NoStop}%
\bibitem [{\citenamefont {Pittalis}\ \emph {et~al.}(2011)\citenamefont
  {Pittalis}, \citenamefont {Proetto}, \citenamefont {Floris}, \citenamefont
  {Sanna}, \citenamefont {Bersier}, \citenamefont {Burke},\ and\ \citenamefont
  {Gross}}]{PPFS11}%
  \BibitemOpen
  \bibfield  {author} {\bibinfo {author} {\bibfnamefont {S.}~\bibnamefont
  {Pittalis}}, \bibinfo {author} {\bibfnamefont {C.~R.}\ \bibnamefont
  {Proetto}}, \bibinfo {author} {\bibfnamefont {A.}~\bibnamefont {Floris}},
  \bibinfo {author} {\bibfnamefont {A.}~\bibnamefont {Sanna}}, \bibinfo
  {author} {\bibfnamefont {C.}~\bibnamefont {Bersier}}, \bibinfo {author}
  {\bibfnamefont {K.}~\bibnamefont {Burke}}, \ and\ \bibinfo {author}
  {\bibfnamefont {E.~K.~U.}\ \bibnamefont {Gross}},\ }\href@noop {} {\bibfield
  {journal} {\bibinfo  {journal} {Phys. Rev. Lett.}\ }\textbf {\bibinfo
  {volume} {107}},\ \bibinfo {pages} {163001} (\bibinfo {year}
  {2011})}\BibitemShut {NoStop}%
\bibitem [{\citenamefont {Dreizler}\ and\ \citenamefont {Gross}(1990)}]{DG90}%
  \BibitemOpen
  \bibfield  {author} {\bibinfo {author} {\bibfnamefont {R.~M.}\ \bibnamefont
  {Dreizler}}\ and\ \bibinfo {author} {\bibfnamefont {E.~K.~U.}\ \bibnamefont
  {Gross}},\ }\href@noop {} {\emph {\bibinfo {title} {Density Functional
  Theory: An Approach to the Quantum Many-Body Problem}}}\ (\bibinfo
  {publisher} {Springer--Verlag},\ \bibinfo {address} {Berlin},\ \bibinfo
  {year} {1990})\BibitemShut {NoStop}%
\bibitem [{\citenamefont {Bartel}\ \emph {et~al.}(1985)\citenamefont {Bartel},
  \citenamefont {Brack},\ and\ \citenamefont {Durand}}]{BBD85}%
  \BibitemOpen
  \bibfield  {author} {\bibinfo {author} {\bibfnamefont {J.}~\bibnamefont
  {Bartel}}, \bibinfo {author} {\bibfnamefont {M.}~\bibnamefont {Brack}}, \
  and\ \bibinfo {author} {\bibfnamefont {M.}~\bibnamefont {Durand}},\ }\href
  {\doibase http://dx.doi.org/10.1016/0375-9474(85)90071-5} {\bibfield
  {journal} {\bibinfo  {journal} {Nuclear Physics A}\ }\textbf {\bibinfo
  {volume} {445}},\ \bibinfo {pages} {263 } (\bibinfo {year}
  {1985})}\BibitemShut {NoStop}%
\bibitem [{\citenamefont {Golden}(1960)}]{G60}%
  \BibitemOpen
  \bibfield  {author} {\bibinfo {author} {\bibfnamefont {S.}~\bibnamefont
  {Golden}},\ }\href {\doibase 10.1103/RevModPhys.32.322} {\bibfield  {journal}
  {\bibinfo  {journal} {Rev. Mod. Phys.}\ }\textbf {\bibinfo {volume} {32}},\
  \bibinfo {pages} {322} (\bibinfo {year} {1960})}\BibitemShut {NoStop}%
\bibitem [{\citenamefont {Cangi}\ \emph {et~al.}(2014)\citenamefont {Cangi},
  \citenamefont {Sim},\ and\ \citenamefont {Burke}}]{CSB14}%
  \BibitemOpen
  \bibfield  {author} {\bibinfo {author} {\bibfnamefont {A.}~\bibnamefont
  {Cangi}}, \bibinfo {author} {\bibfnamefont {E.}~\bibnamefont {Sim}}, \ and\
  \bibinfo {author} {\bibfnamefont {K.}~\bibnamefont {Burke}},\ }\href@noop {}
  {} (\bibinfo {year} {2014}),\ \bibinfo {note} {in preparation}\BibitemShut
  {NoStop}%
\bibitem [{\citenamefont {Roccia}\ and\ \citenamefont {Brack}(2008)}]{RB08}%
  \BibitemOpen
  \bibfield  {author} {\bibinfo {author} {\bibfnamefont {J.}~\bibnamefont
  {Roccia}}\ and\ \bibinfo {author} {\bibfnamefont {M.}~\bibnamefont {Brack}},\
  }\href {\doibase 10.1103/PhysRevLett.100.200408} {\bibfield  {journal}
  {\bibinfo  {journal} {Phys. Rev. Lett.}\ }\textbf {\bibinfo {volume} {100}},\
  \bibinfo {pages} {200408} (\bibinfo {year} {2008})}\BibitemShut {NoStop}%
\bibitem [{\citenamefont {Holmes}(2013)}]{H13}%
  \BibitemOpen
  \bibfield  {author} {\bibinfo {author} {\bibfnamefont {M.~H.}\ \bibnamefont
  {Holmes}},\ }\href@noop {} {\emph {\bibinfo {title} {Introduction to
  Perturbation Methods}}}\ (\bibinfo  {publisher} {Springer},\ \bibinfo {year}
  {2013})\BibitemShut {NoStop}%
\end{thebibliography}%

\end{document}